\documentclass[]{aastex6}

\usepackage{multirow}
\fullcollaborationName{The Friends of AASTeX Collaboration}

\begin{document}
\fontsize{10pt}{10pt}{Comments: Accepted for publication in The Astrophysical Journal. 9 pages, 8 figures, 3 tables.}
\title{Fermi-LAT detection of GeV $\gamma$-ray
emission from Type Ia supernova remnant G272.2-3.2}

\author{Yunchuan Xiang\altaffilmark{1} and Zejun jiang\altaffilmark{1}}

\altaffiltext{1}{Department of Astronomy, Yunnan University, and Key Laboratory of Astroparticle Physics of Yunnan Province, Kunming, 650091, China, zjjiang@ynu.edu.cn, xiang{\_}yunchuan@yeah.net}

\begin{abstract}
A new $\gamma$-ray source with a significance level of approximately 5$\sigma$ was reported in the region of SNR G272.2-3.2, analysing the approximately 12.4 years of observation data from the Fermi Large Area Telescope (Fermi-LAT). Its $\gamma$-ray spatial distribution did not show extended feature, and it had a soft spectrum with the spectral index of 2.56$\pm$0.01 of a power-law model. No significant  variability of its light curve (LC) with 10 time bins was identified, and its spatial positions in the X-ray and GeV bands overlapped. We suggest that the new $\gamma$-ray source is a likely counterpart of SNR G272.2-3.2. Analysing its spectrum, we discussed the likely origins of the $\gamma$-ray emission.
\end{abstract}

\section{Introduction} \label{sec:intro}

A bright X-ray supernova remnant (SNR) was initially reported by  \citet{Pfeffermann1991} and \citet{Aschenbach1993}, using the Roentgen Satellite (ROSAT).  \citet{Greiner1994} subsequently reported on the detailed properties in the X-ray band from this centrally filled SNR with ROSAT and optical observations; they found that a plasma temperature of approximately 10$^{7}$ K and its X-ray emission exhibited a thermal feature. They identified that its center did not have a non-thermal  property. 
\citet{Winkler1993} obtained observation result from the Cerro Tololo Inter-American Observatory; they reported that the nebulosity near the center of SNR G272.2-3.2 had features of SNR filaments, including observed emissions at 732.5 nm and 658.3 nm, and a measured [S II]/H$_{\alpha}$ ratio approximately 1.4 in the optical band. However, there was no significant diffuse emission from the region of SNR G272.2-3.2.

For the lack of radio continuum observations, \citet{Duncan1997} perform detailed  analyses in the radio band for the region of SNR G272.2-3.2, using the Australia Telescope Compact Array, the 64 m Parkes radio telescope, and the Molonglo Observatory Synthesis Telescope. They observed that this SNR had a non-thermal property in the radio band with a spectral index of 0.55 $\pm$ 0.15, and its morphological structure is approximately circular, which is similar to the X-ray morphology. Its diameter is approximately 15 arcmin, and its internal consists of many radio `blobs' from its faint filaments. They suggested that the diffuse emission of the SNR probably originated from shock-accelerated electrons. 
Utilizing the observation results from the Advanced Satellite for Cosmology and Astrophysics and ROSAT satellite,
 \citet{Harrus2001} identified that SNR G272.2-3.2 was a thermal composite SNR according to its thermal emission and non-shell-like morphology features. They found that the property of the X-ray emission  of SNR G272.2-3.2 can be described, using a nonequilibrium ionization model.


\citet{Park2009} reported that the central nebulosity of SNR G272.2-3.2 has a limb-brightened shell in the X-ray energy band, analysing Chandra observations. They found that the X-ray spectra from the shocked interstellar medium region and the outer shell region were consistent, and the stellar ejecta accelerated by the reverse shock were suggested to be from the SNR itself, according to the enhanced abundances of S, Si, and Fe from the central emission of SNR G272.2-3.2.
\citet{Lopez2011} confirmed that SNR G272.2-3.2 is Type Ia SNR, 
because it had homogeneous and symmetric emission and a small P$_{2}$/P$_{0}$ value, where P$_{2}$ and P$_{0}$ are the values of quadrupole and octupole, respectively. 
\citet{Sezer2012} reported elevated abundances of S, Si, Fe, Ca, and
Ni from the central region of SNR G272.2-3.2; these results showed that its X-ray emission originates from its own ejecta. They further compared the best-fit relative abundance ratios of seven elements and the predicted abundance ratios of the delayed detonation and carbon deflagration models \citep{Nomoto1997}; they proposed that G272.2-3.2 is the result of a Type Ia supernova explosion.
\citet{McEntaffer2013} found that its X-ray spectral features were consistent with those of interstellar material heated by shock in a clumpy medium via the Chandra observatory.
Using XMM-Newton and Chandra observations, the elevated abundances of Si, S, and Fe in the central region of this SNR were found, and the results suggested a Type Ia progenitor for this SNR \citep{Sanchez2013}.  
 Moreover, \citet{Kamitsukasa2016} found that the central energy of the Fe K-shell lines (approximately 6.4 keV) was consistent with most of the Type Ia SNRs via the data analysis of Suzaku, which suggests that the reverse shock of this SNR heats ejecta of its interior, and they evaluated the distance range of this SNR to be approximately 2-3 kpc. 
\citet{Leahy2020} estimated age of the SNR to be 7500$^{+3800}_{-3300}$  years, using the SNR models of spherically symmetric evolution.

Diffusive shock acceleration (DSA) is considered to be the predominant mechanism for astrophysical particle acceleration; it can accelerate the energy of cosmic-ray (CR) particles to 100 TeV or higher via SNR shocks  \citep[e.g.,][]{Aharonian2007,Aharonian2011}. Moreover, the potential process of reacceleration from SNRs also can accelerate CR particles to GeV or TeV energy bands \citep[e.g.,][]{Caprioli2018,Cristofari2019}. Therefore, the energy range of (spectral energy distributions) SEDs from SNRs in the Milky Way is likely to reach GeV or higher energy level  \citep[e.g.,][]{Zhang2007,Morlino2012,Tang2013}. 
Thus far, the fourth Fermi catalog (4FGL) contains 24 firmly certified SNRs and 19 SNR candidates \citep{4FGL}.

 No significant GeV $\gamma$-ray emission of the SNR was identified by \citet{Acero2016b}. Its closest TeV source is HESS J0852-463 \citep{HESS2018}. However, there is no convincing evidence to verify  their correlation, owing to a large angular separation of 6.2$^{\circ}$.
Through our preliminary analysis, we found likely GeV $\gamma$-ray radiation in the region of SNR G272.2-3.2 using Fermitools.
This warranted valuable result of the relative GeV properties of the SNR in all subsequent analyses. The remainder of this paper presents the data reduction method in Section 2, details on the source detection in Section 3, and discussion and conclusion about the study in Section 4.

\section{Data Reduction}

We followed the binned likelihood method\footnote{https://fermi.gsfc.nasa.gov/ssc/data/analysis/scitools/binned{\_}likelihood{\_}tutorial.html} for this analysis.
Fermitools with version {\tt v11r5p3}\footnote{http://fermi.gsfc.nasa.gov/ssc/data/analysis/software/} was selected to analyze the region of interest (ROI) of a $20^{\circ}\times 20^{\circ}$ range around the location of SNR G272.2-3.2 (R.A., decl.= 136.71$^{\circ}$, -52.12$^{\circ}$; from SIMBAD\footnote{from http://simbad.u-strasbg.fr/simbad/}).  The instrumental response function ``P8R3{\_}SOURCE{\_}V3''  was selected to analyze these events from the ROI with the commands of  evtype = 3 and evclass = 128. 
To avoid a large point spread function (PSF) in the low-energy band and increases the pollution of diffuse emissions from galactic and extragalactic backgrounds, the photon energy band was selected from 200 MeV to 500 GeV. 
The time range from the Fermi-LAT observations was from August 4, 2008 (mission elapsed time (MET) 239557427) to December 29, 2020 (MET 630970757). 
Photons with the maximum zenith angles $> 90^{\circ}$ were excluded to  suppress the GeV emission contribution from the Earth Limb. The source model file was generated using the script {\tt make4FGLxml.py}\footnote{https://fermi.gsfc.nasa.gov/ssc/data/analysis/user/}, and all sources from the 4FGL\footnote{In the current period, we used the latest 4FGL, gll{\_}psc{\_}v27.fit, for all subsequent analyses. Please refer to https://fermi.gsfc.nasa.gov/ssc/data/access/lat/10yr{\_}catalog/} within 25$^{\circ}$ around the location of SNR G272.2-3.2 were added to the source model file. 
Then we added a point source with a power-law spectrum\footnote{$N(E) = N_{0}(E/E_{0})^{-\Gamma}$, where $\Gamma$ is the spectral index, from https://fermi.gsfc.nasa.gov/ssc/data/analysis/scitools/xml{\_}model{\_}defs.html\#powerlaw} at the SIMBAD location of SNR G272.2-3.2 to the source model file, where 
normalizations and spectral indexes from the 4FGL sources within the 5$^{\circ}$ range of the ROI were set as free, and normalizations of the galactic and extragalactic diffuse backgrounds, including the galactic diffuse emission ({\tt gll{\_}iem{\_}v07.fits}) and the isotropic extragalactic emission ({\tt iso{\_}P8R3{\_}SOURCE{\_}V3{\_}v1.txt}\footnote{please refer to http://fermi.gsfc.nasa.gov/ssc/data/access/lat/BackgroundModels.html}) were also set as free.

\section{Source Detection} \label{sec:data}

\begin{figure}
  \includegraphics[width=60mm,height=60mm]{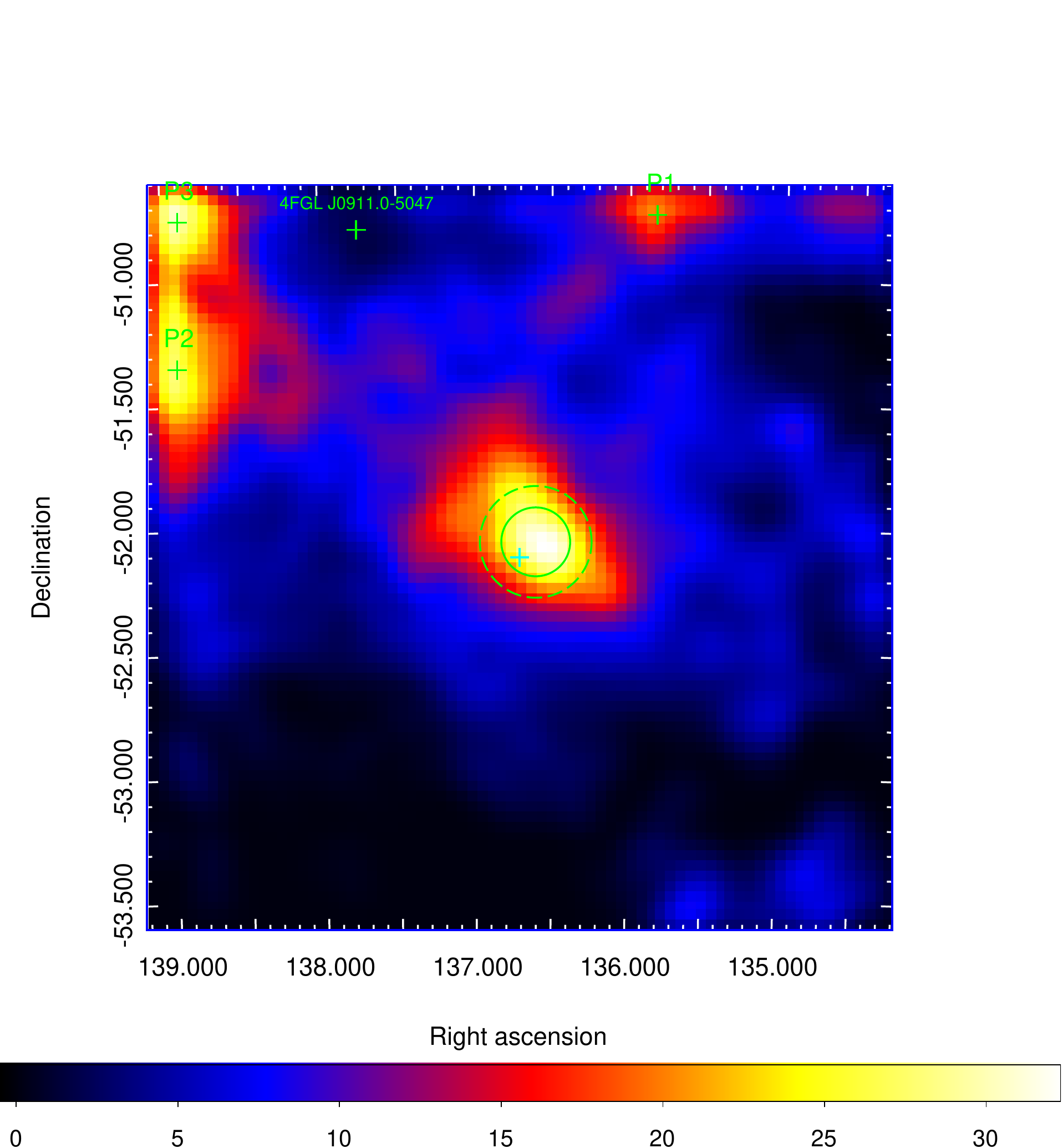}
  \includegraphics[width=60mm,height=60mm]{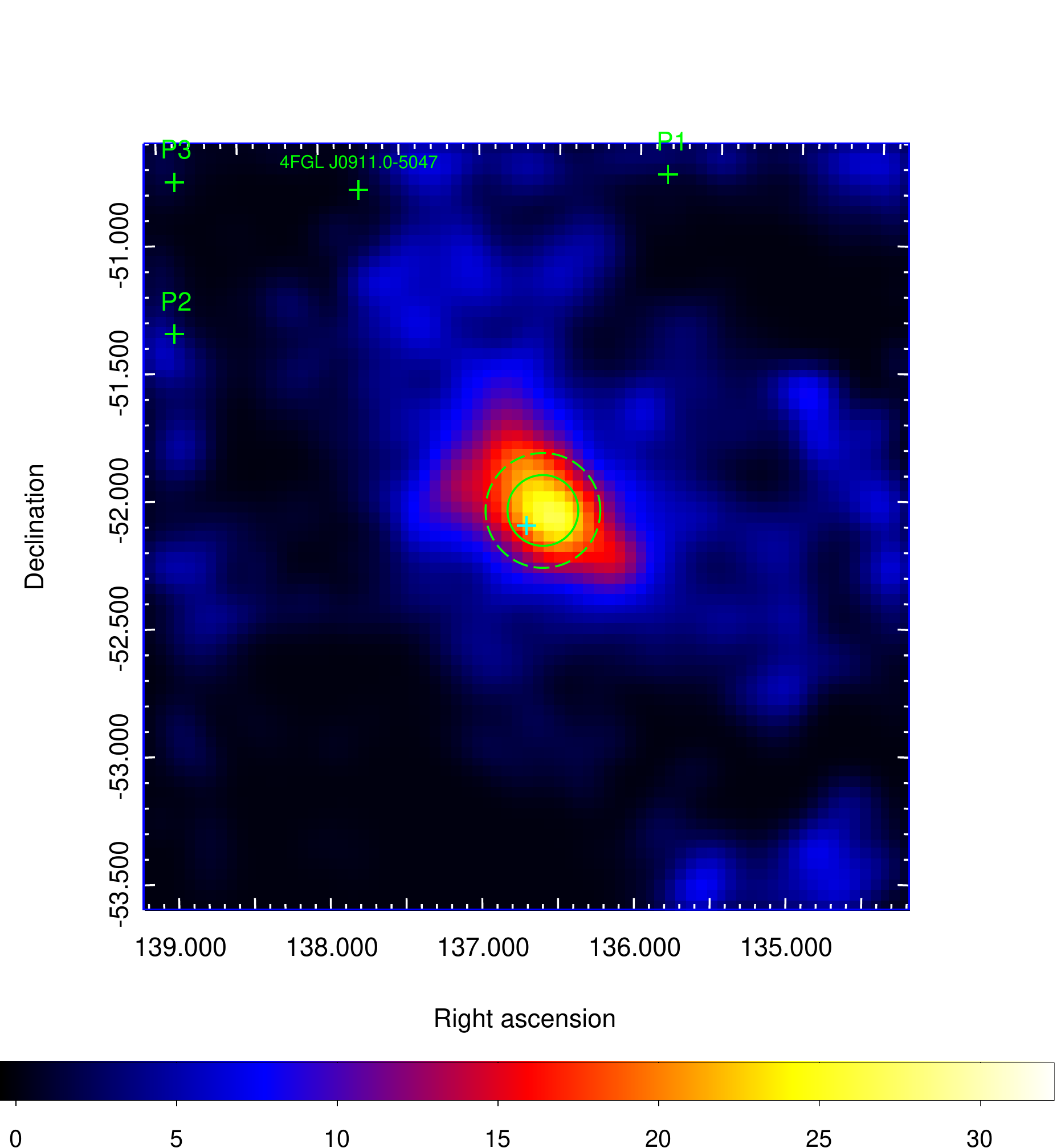}
  \includegraphics[width=60mm,height=60mm]{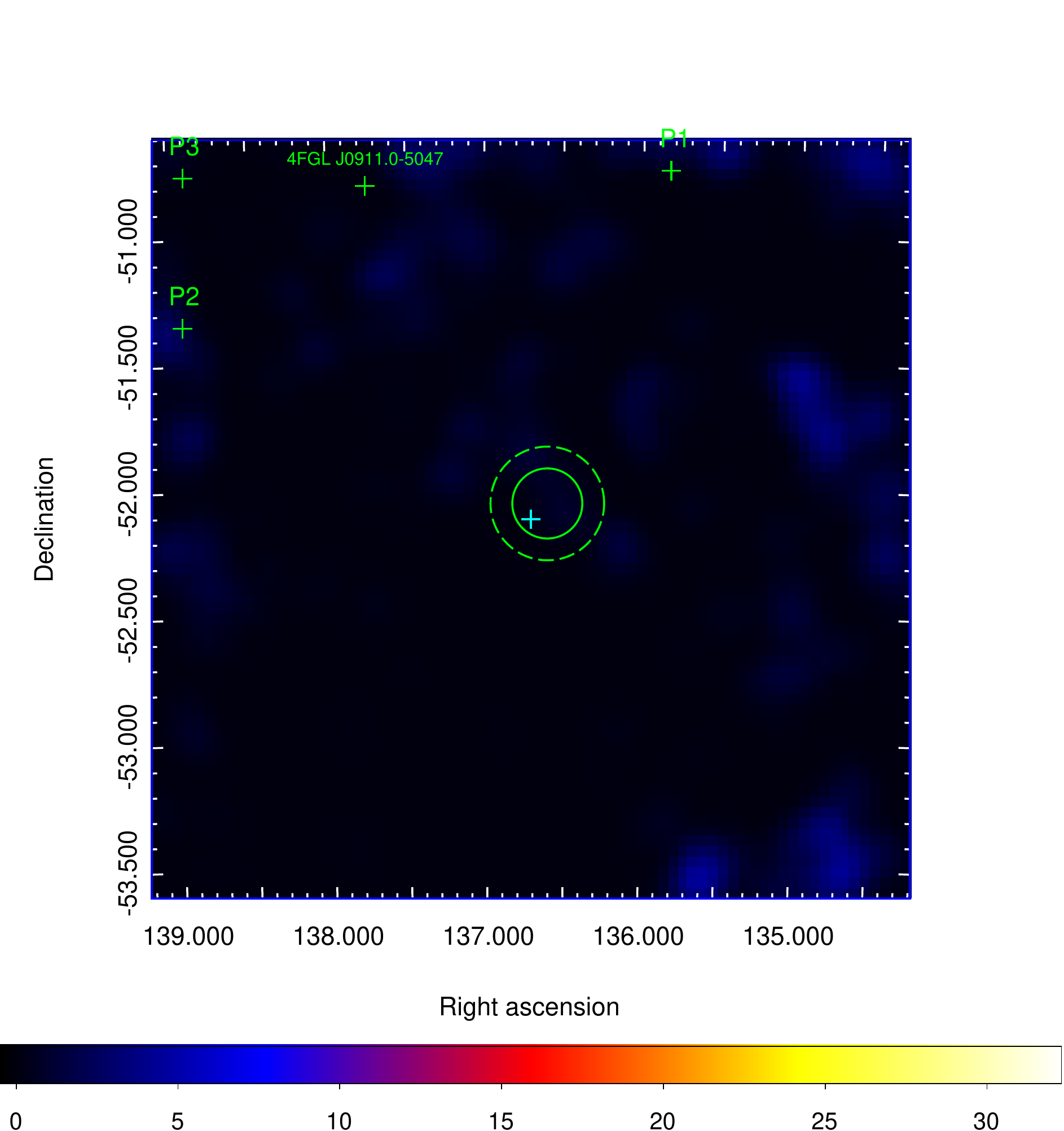}
\caption{
Three $2.6^{\circ}\times2.6^{\circ}$ TS maps with a 0.04$^{\circ}$ pixel size are smoothed using a Gaussian function with $\sigma =0.3^{\circ}$ in the 0.2-500 GeV energy band. Their centers are the location of SNR G272.2-3.2 from SIMBAD, marked as a cyan cross. The 1$\sigma$ (2$\sigma$) error circle of the best-fit location of SNR G272.2-3.2 is represented using a solid (dashed) green circle in the above TS maps. In the left TS map, the background residual radiations are not deducted. 
In the middle TS map, we deducted three significant $\gamma$-ray excesses from the regions of P1, P2, and P3. In the right TS map, we deducted all sources marked in this TS map.}
    \label{Fig1}
\end{figure}

 The 2.6$^{\circ}$ $\times$2.6$^{\circ}$ of TS maps in the 0.2-500 GeV energy band was first generated for the region of SNR G272.2-3.2 using the command {\tt gttsmap}. 
As shown in the left panel of Figure \ref{Fig1}, we found a significant $\gamma$-ray emission from the region of SNR G272.2-3.2  with a TS value of 27.96.
 To exclude significant $\gamma$-ray excesses around the SIMBAD location of SNR G272.2-3.2, three point sources with the power-law spectra, denoted as P1, P2, and P3, were added to three locations of the local maxima of the TS map in all subsequent analyses. Their best-fit results were shown in Table \ref{Table1}.

 \begin{table*}[!h]
\caption{The Best-fit Parameters of the $\gamma$-ray Excesses of P1-P3 in the 0.2-500 GeV Energy Band}
\begin{center}
\begin{tabular}{ccccccc}
    \hline\noalign{\smallskip}
 Source  Name   & $\rm TS$    &  $R.A.$  &  $Decl.$  & $N_{0}$ & $\Gamma$   \\
                    &         &   (deg)  & (deg)     & ($10^{-13}$) &  \\
  \hline\noalign{\smallskip}
         P1    &    19.94  & 135.83 & -50.74 & 0.77 $\pm$ 0.02 &   2.50 $\pm$ 0.01 \\
         P2    &    9.34  & 138.92 & -51.35 & 0.59 $\pm$ 0.17 & 2.51 $\pm$ 0.04 \\     
         P3    &    58.38  & 138.89 & -50.75  & 0.75 $\pm$ 0.01 & 3.11 $\pm$ 0.01 \\ 
  \noalign{\smallskip}\hline
\end{tabular}
\end{center}
\textbf{Note}:
The fitting results of P1-P3 with a power-law spectrum, $dN/dE=N_{0}(E/E_{0})^{-\Gamma}$, $E_{\rm 0}=2$ GeV.
\label{Table1}
\end{table*}
 
 We found that the significant $\gamma$-ray radiation with a TS value of $22.37$ still exists in the region of SNR G272.2-3.2, as shown in the middle panel of Figure \ref{Fig1}. 
To firmly confirm the existence of significant $\gamma$-ray excesses in the SIMBAD location of SNR G272.2-3.2, the radiation from the region of  SNR G272.2-3.2 was also subsequently excluded. No significant $\gamma$-ray excesses in its SIMBAD location were found, as shown in the right panel of Figure \ref{Fig1}. This result further supports that the significant $\gamma$-ray signal exists in the region of SNR G272.2-3.2.

\begin{figure}[!h]
\centering
  \includegraphics[width=\textwidth, angle=0,width=100mm,height=100mm]{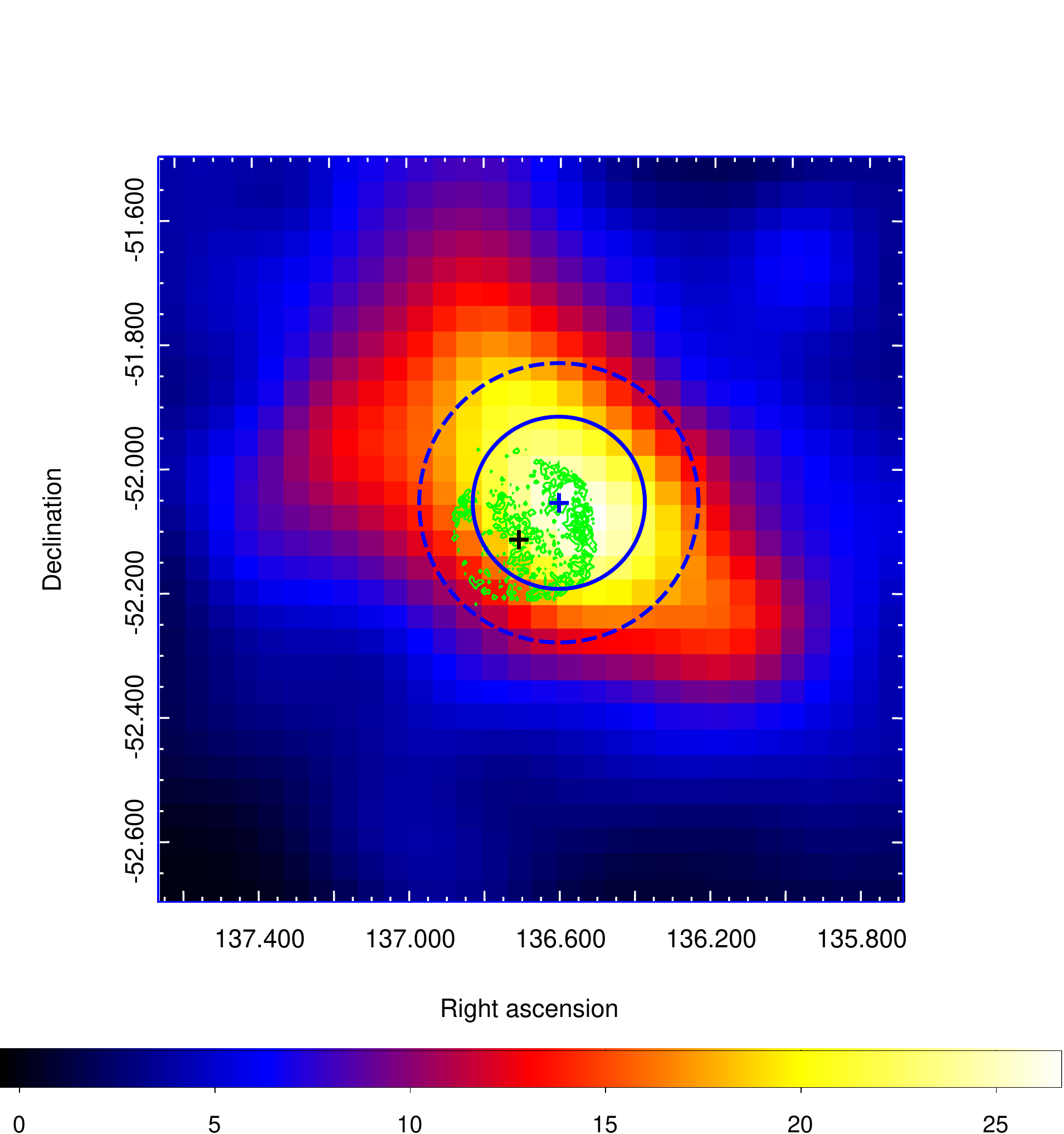} %
 \caption{The $1^{\circ} \times 1^{\circ}$ of TS map with a $0.04^{\circ}$ pixel size in the 0.2-500 GeV band is smoothed with a Gaussian function with $\sigma =0.3^{\circ}$,  and the SIMBAD location of SNR G272.2-3.2 is its center, marked as a black cross. The blue cross is its best-fit location. The solid and dashed blue circles were the 1$\sigma$ and 2$\sigma$ error circles of the best-fit location of SNR G272.2-3.2, respectively. Green contours are from the  observation of XMM-Newton \citep{Sanchez2013}.
 }
 \label{Fig2}
\end{figure}

Additionally, we calculated its luminosity to be (2.66$\pm$0.05)$\times 10^{33} $ $\rm erg\ s^{-1}$ in the 0.2-500 GeV energy band, using a distance of 2.5 kpc from \citet{Kamitsukasa2016}, and the magnitude range  of its luminosity is within those of 10$^{33}$-10$^{35}$ $\rm erg\ s^{-1}$ of the thermal composite SNRs observed in the Milky Way \citep{Liu2015}.
This result implies that the GeV emission may originate in the region of SNR G272.2-3.2. 
The best-fit position (R.A., decl. = 136.60$^{\circ}$, -52.06$^{\circ}$) from the region of SNR G272.2-3.2  was calculated by using \textbf{the command} {\tt gtfindsrc}, and its 
 68\% (1$\sigma$) and 95\% (2$\sigma$) error circles were 0.14$^{\circ}$ and 0.22$^{\circ}$, respectively. 
This result showes that the SIMBAD location of SNR G272.2-3.2 is within the 1$\sigma$ error circle of the best-fit position; most of X-ray contours of the XMM-Newton from the region of  SNR G272.2-3.2 \citep{Sanchez2013} are within the 1$\sigma$ error circle, as shown in Figure \ref{Fig2}. The overlap of the spatial locations of the X-ray and GeV energy bands suggests that the GeV $\gamma$-ray emission is likely to be from SNR G272.2-3.2. Later, the best-fit position of SNR G272.2-3.2  was selected to replace its SIMBAD position for all subsequent analyses.

\setlength{\belowcaptionskip}{0.3cm}

We subsequently tested whether the $\gamma$-ray emission in the 1-500 GeV energy range from the region of SNR G272.2-3.2 has an the extended spatial distribution, using the two-dimensional (2D) Gaussian and uniform disk templates. The different values of the $\sigma$ and radius, which range from 0.05$^{\circ}$ to 2$^{\circ}$ with an increment of 0.05$^{\circ}$, were tested for the two spatial templates. 
We calculated the value of TS$_{\rm ext}=$2log($L_{\rm ext}$/$L_{\rm ps}$), from \citet{Lande2012}, where the maximum log-likelihood values of an extended source and point source are represented by using $L_{\rm ext}$ and $L_{\rm ps}$, respectively.  
The results showed that the TS$_{\rm ext} \approx 0$ for these two spatial templates, 
which suggests that there is no extended feature for the GeV emission of  the region of SNR G272.2-3.2. Remarkably, there are 16 SNRs' candidates that do not has an extended $\gamma$-ray spatial distribution in the 4FGL thus far.
 Here we continued to use the point-source template for all subsequent analyses.

\subsection{\rm Spectral Analysis}
Using the binned likelihood analysis method, the photon flux of the global fit of SNR G272.2-3.2 was calculated to be $\rm (4.04\pm0.10)\times 10^{-9} ph$ $\rm cm^{-2} s^{-1}$ with the spectral index of $2.56\pm0.01$ in the 0.2-500 GeV energy band. 
Investigating the 4FGL, we found that three spectral indexes of the SNR candidates\footnote{The spectral index of SNR G344.7-0.1 is 2.46$\pm$0,09; that of SNR G166.0+4.3 is 2.59$\pm$0.07; that of MSH 17-39 is 2.53$\pm$0.06.} are close to 2.56, which implies that this soft spectrum is likely for SNRs in the Milky Way.

Here,  three frequently-used spectral models in the 4FGL including PowerLaw (PL),  LogParabola (LOG), and PLSuperExpCutoff2 (PLE2)\footnote{https://fermi.gsfc.nasa.gov/ssc/data/analysis/scitools/source{\_}models.html} were selected to do the global fit in the 0.2-500 GeV band with the same binned likelihood analysis method above. Next, we calculated the value of $\rm TS_{curve}$\footnote{$\rm TS_{curve}$ is defined as 2(log $L$(curved spectrum) - log $L$(powerlaw)), and $\rm TS_{curve}>$ 16 suggests a spectrum has significant curve in \citet{Nolan2012}.}, and 
the results from the two curved models of LOG and PLE2 are  approximately 0, which indicates that the spectrum of SNR G272.2-3.2 does not have significantly curve.  This is consistent with most of SNRs in the Fermi Galactic Extended Source catalog (FGES) from \citet{Ackermann2017}. 
 Therefore, we selected the PL spectrum for all subsequent analyses.

\begin{figure}[!h]
  \centering
  \includegraphics[width=110mm]{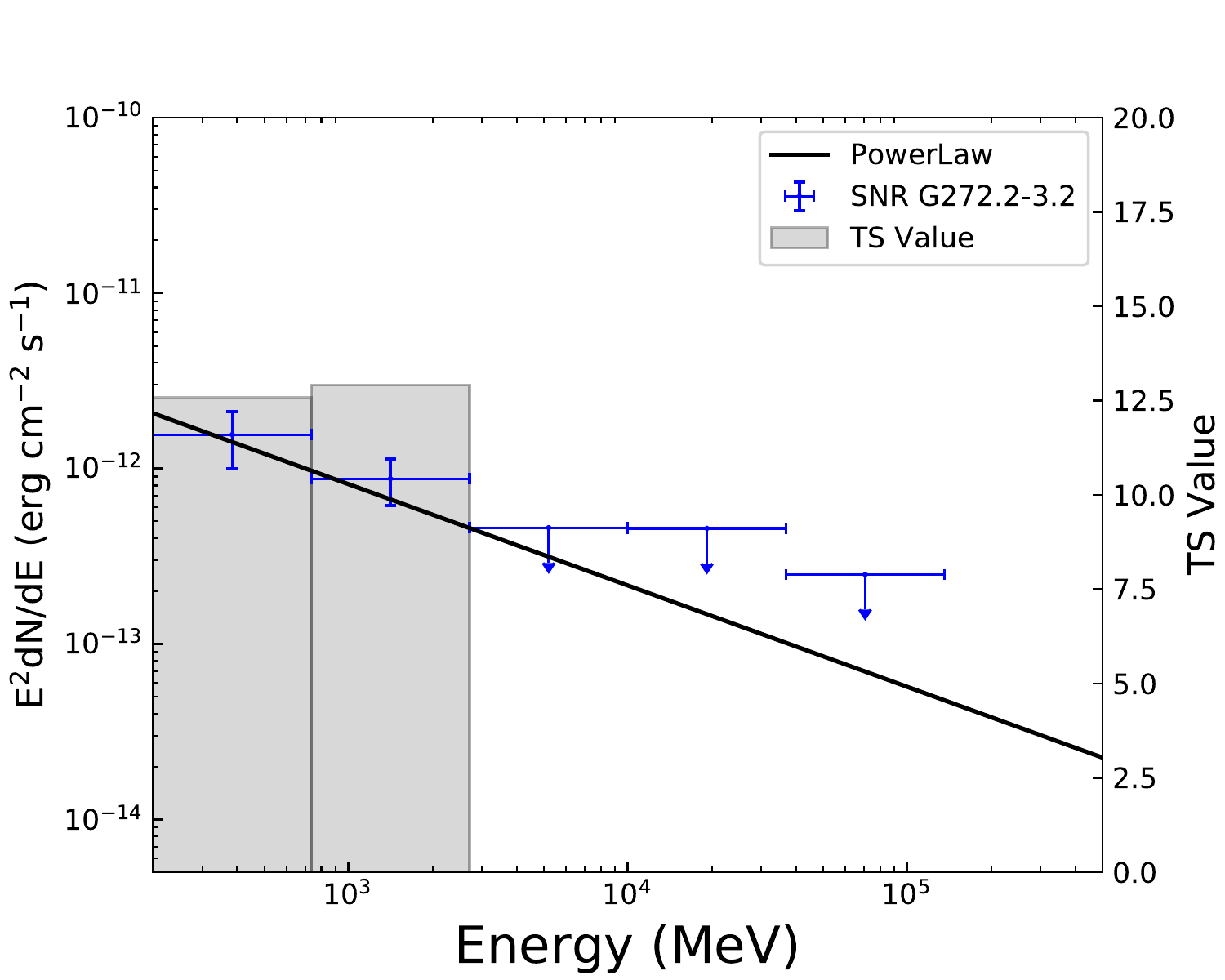}
 \flushleft
\caption{
The SED with five energy bins in the 0.2-500 GeV energy band. 
Blue data points with the total uncertainty including statistical and systematic ones are from this study. The best-fit result of the global fit is represented by using a black solid line.
 The TS values are represented by the gray shaded areas for the energy bins with TS value $>$ 4. The upper limits for the energy bins with TS value $<$4 are calculated.
}
\label{Fig3}
\end{figure}

The SED was generated in the 0.2-500 GeV energy band for SNR G272.2-3.2.
We divided this SED into six equal logarithmic energy bins, and each energy bin was separately fitted using the binned likelihood method.
Additionally, the systematic uncertainty from the effective area was calculated, using the bracketing Aeff method\footnote{https://fermi.gsfc.nasa.gov/ssc/data/analysis/scitools/Aeff{\_}Systematics.html} for the first two bins of the SED.
Upper limits with a 95\% confidence level were provided for the energy bins with TS value $<$ 4. 
The TS values from the fifth and sixth energy bins were all approximately 0;   we only provided the upper limit of the fifth energy bin for the SED, as shown in Figure \ref{Fig3}. The related data from the five energy bins are given in Table \ref{Table2}.

\begin{table}[!h]
\begin{center}
\caption{ The Photon Energy Elux of Each Bin from the SED of SNR G272.2-3.2 Using Fermi-LAT}
\begin{tabular}{lclclclclc}
    \hline\noalign{\smallskip} 
    E  & Band       & $E^{2}dN(E)/dE$ & TS  \\
 (GeV) &  (GeV)     & ($10^{-13}$erg cm$^{2}s^{-1}$ ) &        \\            
  \hline\noalign{\smallskip}
    0.38    & 0.2-0.74    &  15.58$\pm$5.48$^{+0.74}_{-0.82}$ &   12.58  \\
    1.41    & 0.74-2.71    &  8.74$\pm$2.53$^{+0.41}_{-0.46}$  &  12.92      \\
    5.21  & 2.71-10.0    &  4.59  &  2.16   \\
    19.19 & 10.0-36.84   &  4.55  &  1.47  \\
    70.71 & 36.84-135.72 &  2.49  &  0.0   \\     
  \noalign{\smallskip}\hline   
\end{tabular}
    \label{Table2}
\end{center}
\textbf{Note}: 
For the first two energy bins, the first and the second uncertainties represent the statistical and systematic uncertainties, respectively. For other energy bins with TS values $<$ 4, only the 95\% upper limits are only given. 
\end{table}

\subsection{Variability Analysis} \label{sec:data-results}

In the 4FGL, we calculated the average value of the variability index $\rm TS_{var}\approx 10.58<$ 18.48\footnote{A source of $\rm TS_{\rm var}>$ 18.48 for 12 intervals is a variable source with above 99\% chance. Referring to https://heasarc.gsfc.nasa.gov/W3Browse/fermi/fermilpsc.html} from 25 firmly certified SNRs. This result suggested that most of SNRs in the local universe are stable sources in the GeV energy band. 
To test whether the light curve (LC) of SNR G272.2-3.2 has this stable property, we generated an LC of 10 time bins from the 12.4 years of  period in the 0.2-500 GeV energy band.
For this LC, the source of $\rm TS_{\rm var} \geq$ 21.67 was considered to be a variable source with a 99\% confidence level.
The value of $\rm TS_{var}$, defined by \citet{Nolan2012}, was calculated to be $\geq$ 10.53 with a 1.02$\sigma$ variability significance level. This result suggests that no significant variability arises from the LC.

\begin{figure}[!h]
\centering
 \includegraphics[width=\textwidth, angle=0,width=120mm,height=70mm]{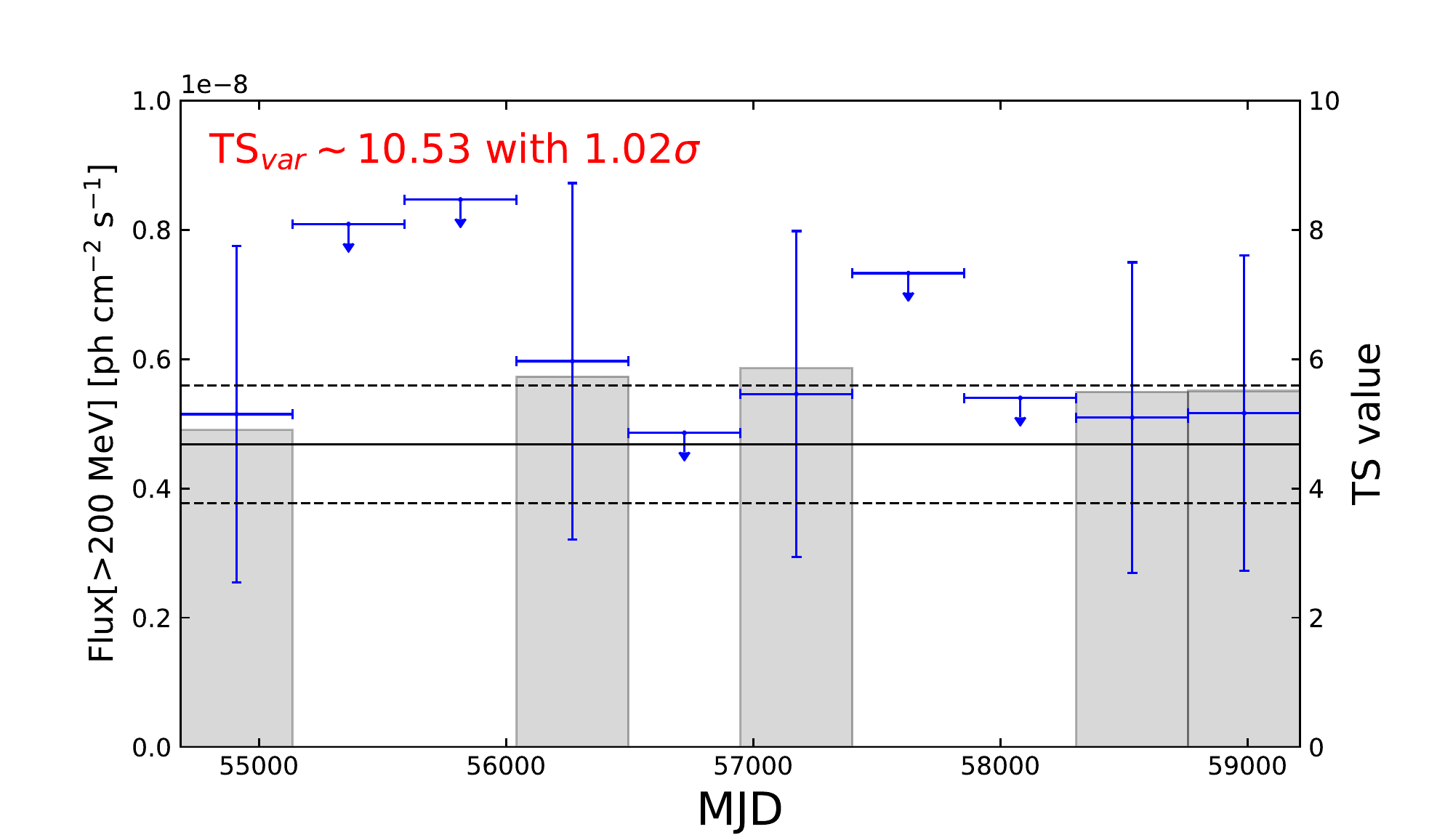} %
 \caption{The panel comprises 12.4 years of the LC with 10 time bins for SNR G272.2-3.2 in the 0.2-500 GeV energy band. 
The upper limits of the 95\% confidence level are calculated for the time bins of TS values $<$ 4. The TS values are represented by gray shaded areas for the time bins with TS values $>$ 4.
The best-fit result from a constant flux model is represented by using a black solid line, and its 1$\sigma$ uncertainties are represented by two black dashed lines.
}
 \label{Fig4}
\end{figure}

To further check the variability from this LC, we also used a constant flux model with the $\chi^{2}$ goodness-of-fit test \cite[e.g., ][]{Zhang2016,Peng2019}.
As shown in Figure \ref{Fig4}, we observed that a simple constant flux model can provide a good fit for data points for a TS value$>$ 1 with a reduced $\chi^{2}$ of 0.12. This further suggests that the LC of SNR G272.2-3.2 has not significant variability of the GeV photon flux thus far.

\section{DISCUSSION and Conclusion} \label{sec:data-results}   

\subsection{\rm Likely origins of the GeV emission of SNR G272.2-3.2}

\begin{figure}[!h]
\centering
 \includegraphics[width=\textwidth, angle=0,width=70mm,height=70mm]{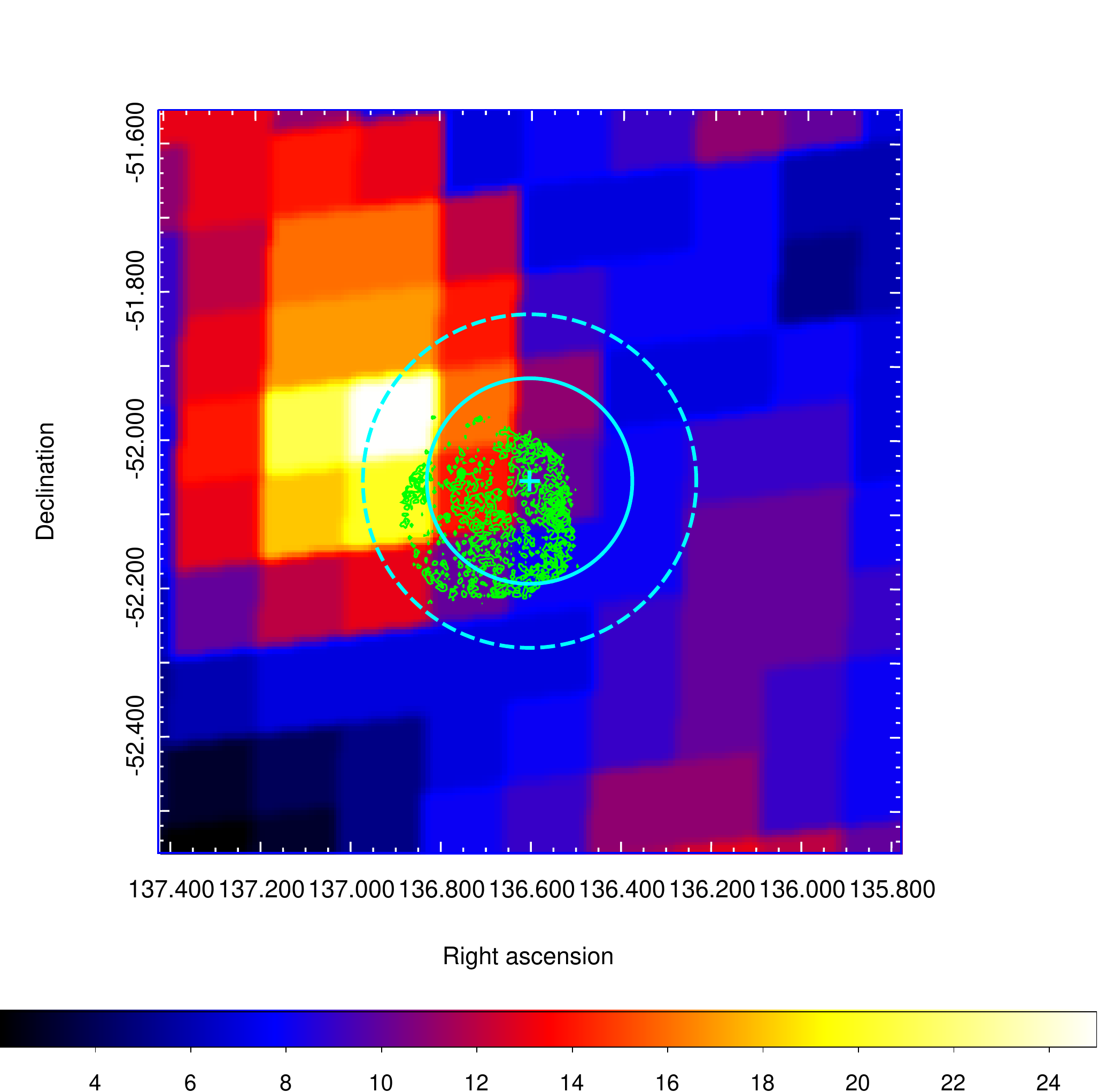} 
 \includegraphics[width=\textwidth, angle=0,width=75mm,height=60mm]{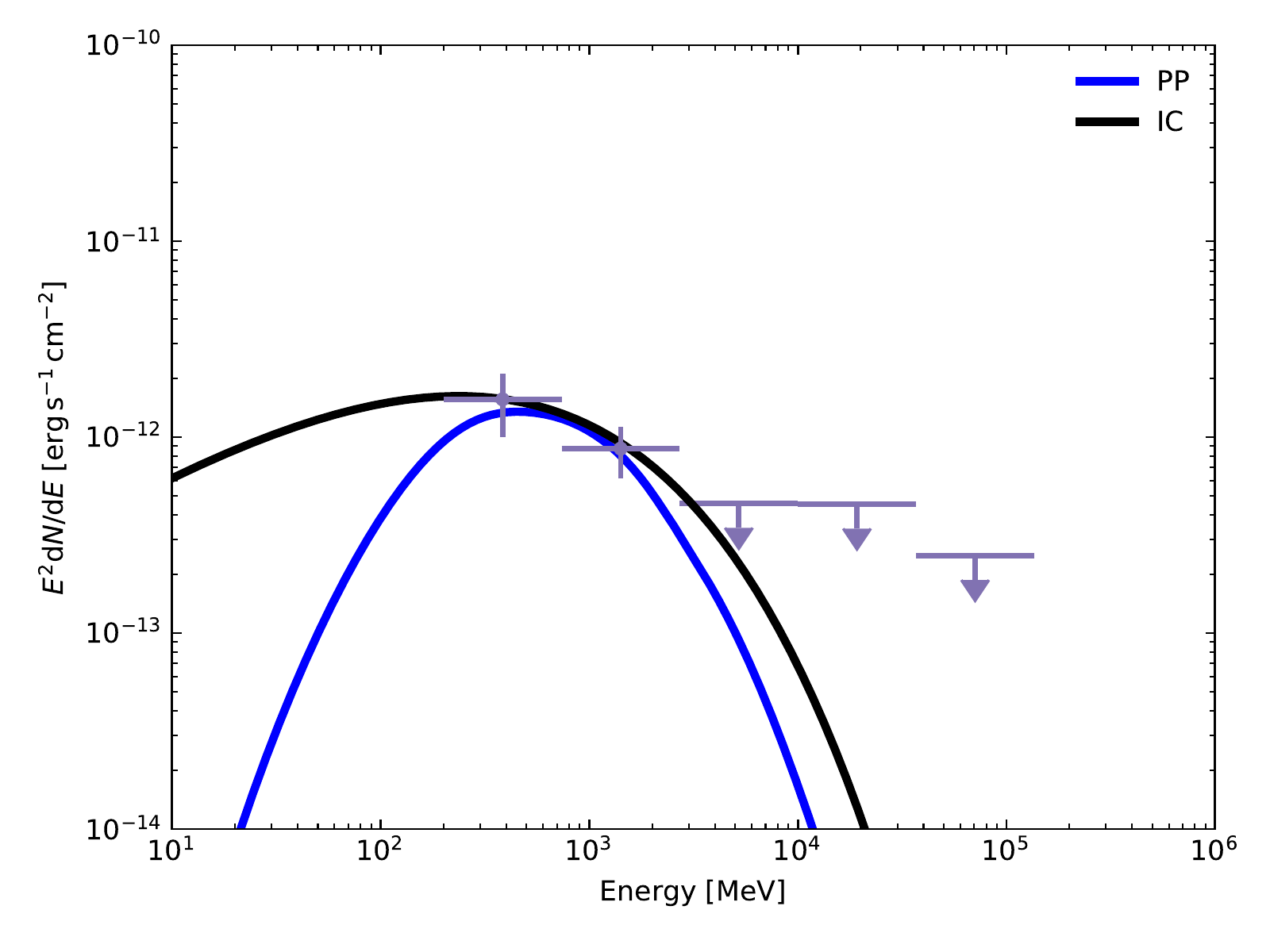}
 \caption{Left panel:  The CO velocity-integrated brightness temperature map from \citet{Dame2001}. The color bar shows CO intensity, and its unite is K km s$^{-1}$.  Two circles and green contours have been described in Figure \ref{Fig2}.  Right panel: The fit result of the SED of SNR G272.2-3.2. The black and blue solid lines represent the fit results of leptonic and hadronic models, respectively.
}
 \label{Fig5}
\end{figure}

Leptonic and hadronic origins are widely used to explain GeV $\gamma$-ray radiation \citep[e.g.,][]{Zeng2017,Zeng2019}. 
The former is generally considered to be caused by the inverse Compton scattering; the latter mainly results from the decay of the neutral pion $\pi^{0}$ from the process in inelastic proton-proton collisions. 
This SNR was considered to be in a middle-aged stage with an age of 7500$^{+3800}_{-3300}$ years \citep{Leahy2020}; 
the GeV $\gamma$-ray emission of a middle-aged SNR is generally considered to be of leptonic origin \citep[e.g.,][]{Guo2017}.
OH maser emission is convincing evidence in verifying a hadronic origin  caused by the interaction of SNR and OH molecular clouds \citep{Frail1996}. However, \citet{McDonnell2008} did not find significant OH maser emissions from this region. 
Significant CO molecular line broadening is also considered as strong evidence, confirming the interactions of CO molecular clouds and relativistic protons \citep[e.g.,][]{Su2017b}. 
We investigated the CO velocity-integrated temperature map from the 1.2 meter millimeter-wave telescope \citep{Dame2001} and found that the CO molecular cloud content was high around this SNR, as shown in the left panel in Figure \ref{Fig5}. Further confirmation of this likely interaction is required (e.g., exploring the broadening of the CO molecular line in the region of SNR G272.2-3.2).

We chose the one-zone model from \textbf{NAIMA} \citep[][and references therein]{Zabalza2015} to explain its SED with leptonic and hadronic scenarios; leptonic and hadronic particle distributions were assumed to be $ N(E) = N_{\rm 0}\left (E/E_{\rm 0} \right)^{-\alpha}\exp\left( -{E/E_{\rm cutoff}}\right)\ $,
 where $N_{\rm 0}$ is the amplitude, $ E_{\rm 0}$ is set to 1 TeV, ${\alpha}$ is the spectral index, $E$ is the particle energy, and $E_{\rm cutoff}$ is the particle cutoff energy \citep{Aharonian2006,Xing2016,Xin2019,Xiang2021}. 
We found that leptonic and hadronic models can explain this observation, as shown in the right panel of Figure \ref{Fig5}.  The fit parameters of the two models are presented in Table \ref{Table3}.

\begin{table*}[!h]
\caption{The Fit Parameters of Leptonic and Hadronic Models}
\begin{center}
\begin{tabular}{lccccccc}
    \hline\noalign{\smallskip}
  Model Name       &  $n_{\rm gas}$    & $\alpha$               & $E_{\rm cutoff}$       & $W_{\rm e}$ (or $W_{\rm p}$) &   Log(Likelihood) & $\chi^{2}/N_{dof}$ \\
                         &  (cm$^{-3}$) &    & (GeV)   & (erg)   & & \\
  \hline\noalign{\smallskip}
  Leptonic model    & --- & 1.58$_{-0.18}^{+0.21}$  & 153.12$_{-35.85}^{+48.35}$ &  2.88$_{-0.65}^{+0.76}\times 10^{48}$ & -0.002 & $\frac{0.002*2}{5-4}=0.004$ \\
      \noalign{\smallskip}\hline
  Hadronic model  & 1  & 2.37$_{-0.15}^{+0.21}$  & 10.62$_{-4.12}^{+6.77}$ &  2.88$_{-0.77}^{+0.79}\times 10^{49}$ & -0.0003 & $\frac{0.0003*2}{5-4}=0.0006$ \\
   \hline\noalign{\smallskip}
\end{tabular}
\end{center}

\label{Table3}
Note: The gas density and distance of SNR G272.2-3.2 are 1 cm$^{-3}$ and 2.5 kpc from \citet{Kamitsukasa2016}, respectively. The leptonic and hadronic energy budgets, $W_{\rm e}$ and $W_{\rm p}$, were calculated $>$ 1 GeV. Log(Likelihood) represents the maximum of the log-likelihood. Three upper limits of SED from this work are included in the fit \citep[e.g.,][]{Abdalla2018}, and $\rm \chi^{2}=-2log(Likelihood)$ \citep{Zabalza2015}.
\end{table*}

We found that the spectral index $\alpha=1.58^{+0.21}_{-0.18}$ is close to the average value for $\alpha=1.64$ from those of shell SNRs with the age of above 1000 years, including RX 1713.7-3946, RX J0852-4622, HESS J1731-347, RCW 86, SN 1006, SNR G150.3+4.5, SNR G296.5+10.0, and SNR G323.7-1.0\footnote{The fitting results of $\alpha$ and $W_{\rm e}$ for RX 1713.7-3946, RX J0852-4622, HESS J1731-347, RCW 86, and SN 1006 from \citet{Acero2015b}. For SNR G323.7-1.0, SNR G296.5+10.0, and SNR G150.3+4.5, we used \textbf{NAIMA} to fit them separately, and their observation data and fitting parameters were selected from \citet{Araya2017}, \citet{Zeng2021} and \citet{Ackermann2017}, respectively. The best-fit results of $\alpha$ and $W_{\rm e}$ for SNR  G150.3+4.5, SNR G296.5+10.0, and SNR G323.7-1.0 are 1.78$^{+0.08}_{-0.12}$ and 3.13$^{+0.70}_{-0.71} \times 10^{46}$ erg, 1.64$^{+0.11}_{-0.14}$ and 4.11$^{+0.72}_{-0.63} \times 10^{47}$ erg, and 1.74$^{+0.11}_{-0.18}$ and 1.50$^{+0.69}_{-0.77} \times 10^{48} $ erg, respectively.}. Its leptonic energy budget $W_{\rm e} \approx 2.88^{+0.76}_{-0.65} \times 10^{48}$ erg is within their energy budget range of $10^{46}$ to $10^{48}$ erg, suggesting that  leptonic origin is likely. The hadronic energy budget $W_{\rm p}$ $\approx$ 2.88$^{+0.79}_{-0.77}$ $\times 10^{49}$ erg and spectral index of 2.37$^{+0.21}_{-0.15}$ are close to those of SNRs with molecular cloud system, including IC 443, W 44, W 51C, W 49B, and Puppis A \citep{Xiang2021}. The results suggest that  a hadronic origin is also likely. Moreover, we  believed that  the coexistence of leptons and hadrons is also likely for SNR G272.2-3.2 \citep[e.g.,][]{Zhang2007,Guo2017,Xin2019,Zeng2019}.

We found that both models had a low particle cutoff energy. Coincidently, the age of SNR G272.2-3.2 likely reaches 11000 years \citep{Leahy2020} and may be in a mid-to-late evolution stage. \citet{Tang2013} modeled the multiwavelength radiation from SNRs, using a time-dependent model; they found that the cutoff energies of electrons and protons gradually  decreased after 2000 years. Afterward, \citet{Zeng2019} also found that the particle cutoff energies gradually decreased with an increase in the SNR age. \citet{Zeng2019} believed that this phenomenon may be caused by a gradually weakening shock with the aging of SNR \citep{Ohira2017,Zhang2017}. 
If SNR G272.2-3.2 is in the Sedov phase, the escape and/or energy loss processes dominate the evolution of high-energy particle distribution, causing the maximum energy and particle cutoff energy to decrease quickly \citep{Helder2012,Ohira2012,Zeng2019}. 
If SNR G272.2-3.2 is in the radiative cooling phase (also known as the snow-plough phase), radiative losses become important in the SNR particle evolution; the velocity of the SNR shock wave continuously slows, and the high-energy particles inside the SNR lose most of their energy through radiative cooling, producing a low particle cutoff energy in the low particle cutoff energy in the SNR \citep[e.g.,][]{Cox1972,Blondin1998,Tang2013,Brantseg2013,Zeng2019,Vink2020}. 
Thus, with a likely age of approximately 11000 years,  SNR G272.2-3.2 probably had a low particle cutoff energy.

\subsection{\rm Summary}
   \begin{enumerate}
      \item A significant $\gamma$-ray emission from the region of SNR G272.2-3.2  with a significance level of approximately 5$\sigma$ was found via the analysis of 12.4 years of Fermi-LAT Pass 8 data in the 0.2-500 GeV energy band. Its photon flux of the global fit is $\rm (4.04\pm0.10)\times 10^{-9} ph$ $\rm cm^{-2} s^{-1}$, and its spectral feature with a soft spectral index of $2.56\pm0.01$ has no significant curve. 
      \item The magnitude of its GeV luminosity is within the range of the thermal composite SNRs in the Milky Way.
      \item The GeV spatial position of SNR G272.2-3.2 is in good agreement with that of the X-ray band from XMM-Newton with a angular resolution of 6$^{''}$\footnote{https://www.cosmos.esa.int/web/xmm-newton/technical-details-epic}.
      \item No extended $\gamma$-ray spatial distribution was found in the region of SNR G272.2-3.2, with TS$_{\rm ext}\approx$0, when the uniform disk and 2D Gaussian spatial templates were checked. 
      \item No significant variability was found in the 12.4 yr period LC, which is in good agreement with most certified GeV SNRs in the 4FGL.
      \item Owing to the above conclusions, we suggest that the new $\gamma$-ray source is a likely counterpart of SNR G272.2-3.2.
      \item The leptonic and hadronic scenarios can explain its SED from this work. The hadronic origin from interactions of this SNR and ambient CO molecular clouds needs to be further confirmed in the future. 
   \end{enumerate}

\section{Acknowledgments} 
We sincerely appreciate the support for this work from
the National Key R\&D Program of China under Grant No. 2018YFA0404204, the National Natural Science Foundation of China (NSFC U1931113, U1738211, 11303012), the Foundations of Yunnan Province (2018IC059, 2018FY001(-003), 2018FB011), the Scientific Research Fund of Yunnan Education Department (2020Y0039).

\end{document}